\documentclass[aps,prb,twocolumn,showpacs,floats,floatfix]{revtex4-1}%
\usepackage{graphicx}
\usepackage{dcolumn}
\usepackage{bm}
\usepackage{amssymb}
\usepackage{amsmath}
\usepackage{amsfonts}%
\providecommand{\U}[1]{\protect\rule{.1in}{.1in}}
\providecommand{\U}[1]{\protect\rule{.1in}{.1in}}
\begin{document}
\title{Superconducting circuit simulator of Bose-Hubbard model with a flat band}
\author{Xiu-Hao Deng}
\author{Chen-Yen Lai}
\email{clai24@ucmerced.edu}
\author{Chih-Chun Chien}
\affiliation{School of Natural Sciences, University of California, Merced, CA 95343, USA.}

\begin{abstract}
Recent advance in quantum simulations of interacting photons using
superconducting circuits offers opportunities for investigating the
Bose-Hubbard model in various geometries with hopping coefficients and
self-interactions tuned to both signs. Here we investigate phenomena related
to localized states associated with a flat-band supported by the saw-tooth
geometry. A localization-delocalization transition emerges in the
non-interacting regime as the sign of hopping coefficient is changed. In the
presence of interactions, patterns of localized states approach a uniform density distribution for repulsive interactions
while interesting localized density patterns can arise in strongly attractive regime.
The density patterns indicate the underlying inhomogeneity of the simulator.
Two-particle correlations can further distinguish the nature of the localized
states in attractive and repulsive interaction regimes. We also survey
possible experimental implementations of the simulator.

\end{abstract}

\maketitle
\affiliation{School of Natural Sciences, University of California, Merced, CA 95343, USA.}


\section{Introduction}

Over the past decade intensive research has been focused on studying strongly
correlated states of interacting photons in lattices using various quantum
systems as simulators~\cite{georgescu2014quantum,
aspuru2012photonic,devoret2013superconducting,houck2012chip,
bruder2005bose,schmidt2013circuit, tanese2014fractal}. Among various
simulation schemes, superconducting quantum circuits have became a
particularly promising platform with a close relation to quantum
computation~\cite{devoret2013superconducting,houck2012chip,
schmidt2013circuit}. Many creative ways to simulate quantum systems with
superconducting circuits have been proposed or implemented. For instance, a
variational optimization over continuous matrix product states has been
implemented by using an open circuit quantum electrodynamics (QED) system to
simulate the ground state of the Lieb-Liniger
model~\cite{barrett2013simulating, eichler2015exploring}. Moreover, a digital
quantum simulator can be realized using qubits in a digital quantum
computation platform, where the corresponding quantum operators of the
simulated system are encoded in the Pauli operators of single qubits along
with a series of one- or two-qubit gates. A simulator for the Fermi Hubbard
model has been demonstrated using a programmable X-mon array and a simulator
of the Bose-Hubbard model using a similar system has been proposed as
well~\cite{Barends:2015tv}.

On the other hand, an analog quantum simulator can provide an intuitive
description of many-body systems. One class of simulators can be built with an
array of superconducting circuit elements usually fabricated on a
chip~\cite{buluta2009quantum,
devoret2013superconducting,houck2012chip,schmidt2013circuit}. The quanta of
the excitations on those circuit elements simulate an ensemble of quantum
particles. For example, coupled superconducting qubits as an analogue of a
spin array can be a simulator of the Ising
model~\cite{kadowaki1998quantum,zhang2014quantum}. Alternatively, photonic
excitations in a circuit-QED array may serve as an analogue of lattice
bosons~\cite{blais2004cavity, lang2011observation,hoffman2011dispersive}, and
effective photon interactions could be created by utilizing strong
light-matter couplings between superconducting resonators and
qubits~\cite{blais2004cavity, lang2011observation,
hoffman2011dispersive,devoret2013superconducting,houck2012chip}. By
fabricating circuit QED elements in desired patterns, various lattice
structures can be explored and local controls over coherent or dissipative
dynamics can also be studied~\cite{schmidt2013circuit,deng2015sitewise,
lang2011observation,hoffman2011dispersive,baust2014tunable,bialczak2011fast,chen2014qubit}%
. The latter scheme further allows for simulations of photonic or polaritonic
Bose Hubbard model (BHM)~\cite{schmidt2013circuit, deng2015sitewise}.

The BHM has been an important paradigm in many-body
theories~\cite{fisher1989boson,bloch2012quantum}, in particular the Mott
insulator-superfluid (MI-SF) phase transition it describes. The Mott insulator
is a localized state occurring at integer fillings when interaction energy
dominates, while the superfluid is a delocalized state where kinetic energy
dominates. In superlattices or other complex geometries, the BHM can exhibit
many interesting phases and
phenomena~\cite{buonsante2004topology,roth2003phase,jaksch1998cold,jo2012ultracold,
buonsante2005fractional}. In certain geometries with multiple sites per unit
cell, some of the bands known as flat bands can become non-dispersive.
Particles in a flat band form degenerate localized eigenstates, and this
particular feature may lead to interesting phases including
supersolid~\cite{Huber:2010bc} or topological
insulator~\cite{bergholtz2013topological}. The degeneracy of a flat band, on
the other hand, is very sensitive to external perturbations and can be lifted
easily. The phase diagram of the BHM is enriched if the system supports a flat
band, which can be constructed in several known
geometries~\cite{buonsante2004topology,Flach:2014cm,bergholtz2013topological,parameswaran2013fractional}%
. A number of constructions of flat-band lattices using graph theory also have
been
suggested~\cite{derzhko2010low,mielke1991ferromagnetic,tasaki1992ferromagnetism}%
. Some of the geometries supporting flat bands can be realized in quantum
simulators such as optical lattices for cold atoms or photonic crystals using
micro cavities~\cite{zhang2015onedimensional,
bloch2008many,masumoto2012exciton, christodoulides2003discretizing}, although
having broadly tunable parameters and periodic boundary conditions remains a
great challenge.

Motivated by great opportunities from superconducting circuit simulators, we
investigate the BHM on the saw-tooth lattice using an array of superconducting
circuit elements with tunable couplers. From the analysis of a superconducting
circuit simulator of the BHM outlined in Ref.~\onlinecite{deng2015sitewise},
the simulator in the dispersive regime has a widely tunable parameter range
according to its architecture. The superfluid (SF), Mott insulator (MI), and
the MI-SF transition may be demonstrated and
manipulated~\cite{deng2015sitewise}. A recent experiment has shown
possibilities of simulating attractive bosons modeled by the BHM using an
array of transmons~\cite{hacohen2015cooling}. Due to intrinsic anharmonicity
of transmons, it is challenging for the simulator of
Ref.~\onlinecite{hacohen2015cooling} to exhibit a MI-SF transition or
investigate the repulsive regime. A simulator capable of exploring the BHM
with attractive as well as repulsive interactions, positive as well as
negative hopping coefficients, and flexible geometry allowing for a flat
band~\cite{Flach:2014cm} will elucidate rich physics of the BHM. In the
following we will outline a simulator based on
Ref.~\onlinecite{deng2015sitewise} that can simulate non-interacting photons
and photons with repulsive or attractive onsite interactions with positive or
negative photon hopping coefficients. Interesting localization phenomena
associated with flat bands in selected geometries and energy competitions in strongly interacting regimes can be demonstrated by the
simulator with available experimental parameters. Moreover, the interesting localization and delocalization phenomena survive in the presence of small
fluctuations of system parameters.

This paper is organized as follows. Sec.~\ref{sec:bhm} introduces the Bose
Hubbard model on the saw-tooth lattice and flat-band physics in the
thermodynamic limit. In Sec.~\ref{sec:circuit}, we review the superconducting
circuit simulator for the BHM on periodic saw-tooth lattices and its
experimental parameter range. Sec.~\ref{sec:circuit} presents a discussion on
a localization-delocalization transition due to the presence of a flat-band
and its experimental signatures. Moreover, different ground states in the
repulsive and attractive regimes can be distinguished by two-particle
correlation functions. Relevant issues on experimental realizations of the
simulator are also discussed. Finally, Sec.~\ref{sec:conclusion} concludes our work.

\section{Bose Hubbard model on saw-tooth lattice}

\label{sec:bhm} 
\begin{figure}[t]
\begin{center}
\includegraphics[width=0.48\textwidth]{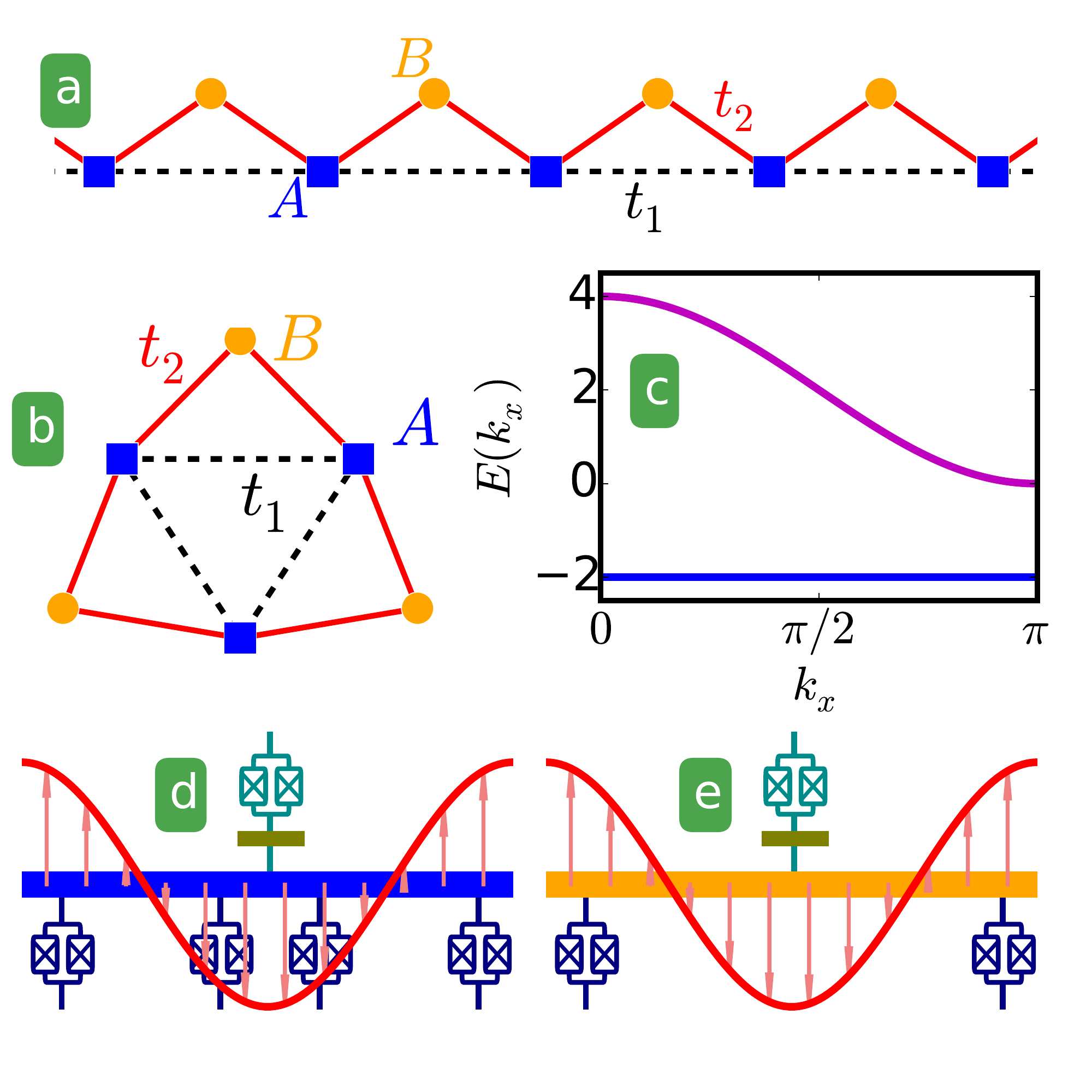}
\end{center}
\caption{(Color online) (a) A segment of the saw-tooth lattice. (b) A periodic
saw tooth lattice. Here the squares (circles) denote the A-sites (B-sites),
and the solid (dashed) lines denote the A-B (A-A) links. (c) Tight-binding
bands of the saw-tooth lattice in the thermodynamic limit with a negative
tunneling coefficient. (d) The elements forming the A-site in (b). (e) The
elements forming the B-site in (b). In (d) and (e), the thick horizontal lines
denote the TLRs and the couplers are made of SQUIDs (with two Josephson
junctions in each loop). Each A-site is connected to four neighbors by the
SQUIDs coupled to the positions depicted in (d). Each B-site is connected to
two neighbors as depicted in (e). The sinusoidal curves illustrate the
fundamental modes in the TLRs. The charge qubits correspond to the SQUIDs
above the TLRs in (d) and (e), and they are coupled to the TLRs via capacitors
(the short horizontal lines above the TLRs) to tune the effective on-site
photon-photon interaction. }%
\label{fig:lattice}%
\end{figure}

In the infinite saw-tooth lattice shown in
Fig.~\ref{fig:lattice}(a) with a particular ratio of hopping coefficients, a
flat band is separated from the other dispersive band as shown in
Fig.~\ref{fig:lattice}(c). Whether the flat band is the lowest or the highest
energy band depends on the sign of hopping coefficients, and interesting
phases can arise due to the flat band. A quantum simulator using
superconducting circuits, capable of demonstrating site-wise manipulations of
the BHM~\cite{deng2015sitewise}, is implemented here to demonstrate signatures
of the flat band structure (FBS). In the following we consider realistic
experimental parameters. Localized states in the flat band are very sensitive
to inhomogeneity of the system, and we will show that the patterns of density
distributions can be used as a probe of imperfections of the simulator.

In the tight-binding approximation, the BHM Hamiltonian on the saw-tooth
lattice is
\begin{align}
\mathcal{H}_{0}  &  =-t_{1}\sum_{\langle ij\rangle}\left(  a_{i}^{\dagger
}a_{j}+a_{j}^{\dagger}a_{i}\right)  -t_{2}\sum_{\langle ij\rangle}\left(
a_{i}^{\dagger}b_{j}+b_{j}^{\dagger}a_{i}\right) \nonumber\\
&  +\sum_{i}U_{i}a_{i}^{\dagger}a_{i}(a_{i}^{\dagger}a_{i}-1)+\sum_{i}%
U_{i}b_{i}^{\dagger}b_{i}(b_{i}^{\dagger}b_{i}-1)\nonumber\\
&  +\sum_{i}\mu_{ai}a_{i}^{\dagger}a_{i}+\sum_{i}\mu_{bi}b_{i}^{\dagger}b_{i}
, \label{eq:ham}%
\end{align}
where $a_{i}(a^{\dagger}_{i})$ and $b_{i}(b^{\dagger}_{i})$ are the
annihilation (creation) operators on the sub-lattice $A$ and $B$ shown in
Fig.~\ref{fig:lattice}(a). To simplify the discussion, we choose $\mu_{i}=0$
across the lattice. When $t_{2}=\sqrt{2}|t_{1}|$ and $U=0$, the lattice
supports a flat band~\cite{Flach:2014cm}, which is the lowest-energy band if
$t_{1} <0$. The Hamiltonian has the energy spectrum
\begin{equation}
E(k_{x})=2t_{1}\text{ and }-2t_{1}(1+\cos k_{x}a)
\end{equation}
depicted in Fig.~\ref{fig:lattice}(c) with a negative value of $t_{1}$. Here
$a$ is the lattice constant and will serve as the length unit. In a finite
periodic lattice, the flat band appears as a set of degenerate localized
states that are eigenstates of the Hamiltonian. Due to a lack of kinetic
energy, flat-band states do not participate in transport. In contrast to the
Mott insulator existing only at integer fillings, the flat-band states are due
to the underlying geometry and could be understood from a single-particle
picture. To make connections with realistic superconducting circuit simulators
consisting of finite numbers of elements~\cite{deng2015sitewise}, we consider
a periodic lattice, for example the one with three unit cells shown in
Fig.~\ref{fig:lattice}(b).

One important feature of the simulator discussed here is that the hopping
coefficients of bosons can be tuned to positive ($t_{1}>0$) or negative
($t_{1}<0$) values. When there is no interaction, the flat band is the lowest
energy band if $t_{1}<0$, so in the ground state the system favors localized
states in the flat band. On the other hand, if $t_{1}>0$, the lowest-energy
band is dispersive and a uniform ground-state density distribution from
delocalized states is expected. Therefore, a localization-delocalization
transition occurs as $t_{1}$ changes sign, which could be realized and
observed in the proposed simulator by tuning the coupler connecting adjacent
lattice sites.

\section{Superconducting circuit simulator}

\label{sec:circuit} Here we briefly review the superconducting circuit
simulator and details can be found in Ref.~\onlinecite{deng2015sitewise}.
Fig.~\ref{fig:lattice}(d)-(e) illustrate the elements and their couplings in
the simulator, which utilizes an array of superconducting transmission line
resonators (TLRs) representing the sites in the BHM. Microwave photons in the
TLRs will simulate the bosons in the BHM \cite{deng2015sitewise}. Adjacent
sites are connected via superconducting quantum interference devices (SQUIDs).
In addition, each TLR is capacitively coupled to a tunable charge qubit with a
SQUID-like structure~\cite{makhlin2001quantum, deng2015sitewise}, which can be
used to tune the effective on-site interaction of the BHM.

The Hamiltonian of the simulator consists of on-site terms and coupling terms
\begin{equation}
H=\sum_{i}\mathcal{H}_{i}^{site}+\sum_{\langle ij\rangle}\mathcal{V}%
_{ij}^{couple}. \label{eq:originsc}%
\end{equation}
Here $\langle ij\rangle$ denotes neighboring pairs as shown in
Fig.~\ref{fig:lattice}(b). The on-site term, $\mathcal{H}_{i}^{site}$, models
a combination of one TLR and a superconducting charge qubit coupled via a
capacitor. In Fig.~\ref{fig:lattice}(d) and (e), a TLR is represented by a
thick horizontal line, the charge qubit is shown as the SQUID above the TLR,
and the capacitor is denoted by a short line in between them. Details of the
modeling are given in Appendix.~\ref{app1}.

A deeply off-resonant qubit in the dispersive regime coupled to a TLR gives
rise to the following Hamiltonian with an effective on-site photon-photon
interaction~\cite{deng2015sitewise}%
\begin{equation}
\mathcal{H}_{i}^{site}=\sum\limits_{i}[\omega_{i}^{\ast}c_{i}^{\dagger}%
c_{i}+U_{i}^{p}c_{i}^{\dagger}c_{i}(c_{i}^{\dagger}c_{i}-1)]. \label{Honsite}%
\end{equation}
Here $\omega_{i}^{\ast}$ is the dressed TLR frequency~\cite{blais2004cavity}.
With rotating wave approximation and dispersive condition, the on-site
repulsion strength $U_{i}^{p}$ can be controlled by qubit-TLR coupling and
detuning as explained in Appendix.~\ref{app1}. By adjusting the detuning
between the qubit and TLR, $U_{i}^{p}$ can be positive or negative.

The coupling term $\mathcal{V}_{ij}^{couple}$ models a coupler SQUID
consisting of two Josephson junctions, which gives rise to a sum of a fixed
capacitive coupling and a tunable inductive coupling between neighboring
sites~\cite{deng2015sitewise,peropadre2013tunable,wulschner2015tunable}.
Explicitly,
\begin{equation}
\mathcal{V}_{ij}^{couple}=-(g^{cap}+g^{ind})(c_{i}^{\dagger}c_{j}+c_{i}%
c_{j}^{\dagger}), \label{eq:rwalink}%
\end{equation}
where $g^{cap}~(g^{ind})$ is the capacitive (inductive) coupling constant
across the link $ij$. Here we assume all the SQUID couplers are identically
fabricated. As shown in Fig.\ref{fig:lattice} (d), the two coupler SQUIDs
below the middle of the TLR are placed at $3/8$ and $5/8$ of the TLR, which
fine tune the ratio of the hopping coefficients between the A-A and A-B links
in the saw-tooth lattice. Furthermore, the inductive coupling constant,
$g^{ind}$, can be tuned to positive or negative values by changing the
magnetic flux through the SQUID.

Here, we limit $\mathcal{V}^{couple}_{ij}$ to the weak coupling regime and
keep only the lowest order when modeling the onsite interaction of
Eq.~\eqref{eq:ham}. The total Hamiltonian now has the Bose-Hubbard form
\begin{align}
H  &  =\sum\limits_{i}[\omega_{i}^{\ast}c_{i}^{\dagger}c_{i}+U_{i}%
c_{i}^{\dagger}c_{i}^{\dagger}(c_{i}c_{i}-1)]\nonumber\\
&  -\sum\limits_{<ij>}(g^{cap}+g^{ind})(c_{i}^{\dagger}c_{j}+c_{i}%
c_{j}^{\dagger}). \label{PerturbedH}%
\end{align}
Compared to Eq.~\eqref{eq:ham}, we can construct a mapping between this
circuit model and BHM where hopping coefficient $t_{i}=g^{cap}+g^{ind}$,
onsite energy $\mu_{i}=\omega_{i}^{\ast}$, and onsite coupling constant
$U_{i}=U_{i}^{p}$. Different values of $t_{1}$ and $t_{2}$ can be obtained by
adjusting $g^{ind}$ and both signs of hopping coefficients can be achieved by
using typical experimental data summarized in Appendix~\ref{app1}. This
feature makes the simulator particularly suitable for studying flat-band
induced phenomena because by changing the sign of $t_{i}$, relative orders of
the energy bands can be reversed. We remark that this circuit model is derived
in the deep dispersive regime, where the on-site qubit is not excited due to a
large detuning. The only on-site excitation quanta are resonant photons
behaving like bosons in the TLR. Therefore, Eq.~\eqref{PerturbedH} describes
the photonic BHM.

By consulting available experimental data (summarized in Appendix~\ref{app1}),
in the following we estimate $t_{i}$ in the range of $-10$MHz to $10$MHz and
sample three regimes in the phase diagram of interacting photons using the
superconducting circuit simulator with uniform $U_{i}=U$: (a) $U\in
\lbrack-5,-0.1]$MHz, (b) $U=0$, and (c) $U\in\lbrack0.1,5]$MHz. For $|U|>5$
MHz, the qubit-TLR detuning $\Delta_{i}$ may be too small and the on-site
excitations could become
polariton-like~\cite{leib2012networks,koch2009superfluid, seo2015quantum}.
Details of how $\Delta_{i}$ is derived can be found in the Appendix. Although
a polaritonic circuit QED lattice may also simulate the BHM with attractive or
repulsive effective interactions, the detailed expressions of the on-site
energy and interaction are different from the photonic model presented here.
The polaritonic system is beyond the scope of our discussion and here we focus
on how the photonic simulator can reveal interesting phases in the BHM when a
flat band is present. In the photonic simulator, it is difficult to approach
the $U=0$ point from finite-$U$ because $U$ depends monotonically on the
detuning $\Delta_{i}$. Thus, opposite signs of $U$ have to be achieved by
starting with opposite signs of the detuning when ground-state behavior is
investigated. To access the $U=0$ point, one may detach the qubit from the TLR
and completely shut down the onsite interaction (see Appendix~\ref{app1} for
more details). Figure~\ref{fig:main} (a) shows finite-$U$ regimes and the
noninteracting regime accessible by the photonic simulator.

\subsection{Inhomogeneity of the simulator}

\label{Ununiform} A realistic simulator will inevitably have imperfections
from its fabrication and operation. As a consequence, $\omega_{i}^{\ast}$ and
$t_{i}$ in each sample will fluctuate rather than being perfectly uniform. In
addition, quantum fluctuations in Josephson junctions can further contribute
to imperfections of superconducting circuit simulators. Here we assume
variations of $\omega_{i}^{\ast}$ and $t_{i}$ due to unwanted cross-talks
between the TLRs can be minimized by carefully designing the chip shielded
from external devices. The fluctuations of $\omega_{i}^{\ast}$ in different
TLRs are estimated as $\delta\omega/\omega=\left\langle \left\vert \omega
_{i}^{\ast}-\omega_{j}^{\ast}\right\vert /\omega_{i}^{\ast}\right\rangle
\sim0.1\%$ based on the following analysis. From available experimental data
showing typical TLR frequencies accurate up to $10^{-3}$ GHz even in a
multi-resonator system~\cite{mariantoni2011photon,hacohen2015cooling}, we
estimate that the inaccuracy of resonator frequency is around $0.1\%$
considering the typical value of the resonant frequency $\sim10$ GHz. The high
quality factors $Q>10^{4}$ of TLRs~\cite{wang2009improving} and photon life
time up to milli-seconds~\cite{reagor2013reaching} ensure that quantum
fluctuations of $\omega_{i}$ are much smaller than fluctuations from
fabrications. Thus, the uncertainties are mainly due to geometrical variation
of the TLRs, which is around $\delta l/l\sim0.1\%$
~\cite{matsuda2012monolithically}. The inaccuracy of resonator frequency due
to variation of the TLR width can be minimized to about $10^{-4}$
GHz~\cite{underwood2012low}. Hence, the variance of resonant frequency
$\omega^{r}\!=\!\frac{2\pi}{\sqrt{C^{r}L^{r}}}$ due to non-uniformity of the
resonator length, which mainly affects the capacitance $C^{r}$, is
approximately $\delta\omega\!/\omega\propto\delta l/l\sim0.1\%$. Furthermore,
the dressed frequency $\omega^{\ast}$ of the TLRs can be finely adjusted by
tuning the qubit energy \cite{blais2004cavity}, which indicates a feasible way
to calibrate the on-site energy $\mu_{i}$. This leads to a even smaller
variance of $\delta\omega\!/\omega$.

For the Josephson junctions in the simulator, there can be more uncertainties
in their fabrications leading to larger deviations of the critical current and
effective capacitance, which in turn cause variation of $t_{i}$ to be
about$1\%$
\cite{shcherbakova2015fabrication,deng2013protecting,mooij1999josephson,wallraff2004strong}%
. Another source of stochastic fluctuations is the magnetic flux noise through
a SQUID exhibiting a typical $1/f$ power spectrum in the range of $1\sim10Hz$
~\cite{lanting2009geometrical}. The flux noise gives rise to variation of the
Josephson energy of SQUIDs and leads to variations of hopping coefficients
$t_{i}$ on the order of $10$kHz~\cite{mooij1999josephson}. Inhomogeneity of
$t_{i}$ due to the noise is estimated as $\delta t/t=\left\langle \left\vert
t_{i}-t_{j}\right\vert /t_{i}\right\rangle \sim10$kHz $/0.1$GHz$=0.01$.
Combining imperfections from fabrication and noise, we estimate $\delta
t/t\sim1\%$, which is about one order of magnitude larger than $\delta
\omega/\omega$. In the following, our numerical simulations will present
different ground-state properties assuming fluctuations are dominated by
$\delta t$. Although there may also be imperfections in $U$, they do not
introduce more physics when $U$ is finite, so $\delta U$ will be neglected.

\section{Results and discussions}

\label{sec:result} A flat band can be seen clearly in energy spectrum in the
thermodynamic limit shown in Fig.~\ref{fig:lattice} (c). However, in a finite
system the boundaries will distort the flat-band and blur its features. To
circumvent finite-size effects, it is desirable to construct a simulator with
periodic boundary condition. Fabrication of superconducting circuits gives the
simulator considered here some advantages when compared to other possible
schemes such as optical lattices \cite{zhang2015onedimensional} where periodic
boundary condition can be difficult. By considering the small number of
elements on superconducting chips in current
experiments~\cite{mariantoni2011photon, houck2012chip, Barends:2015tv}, we
start with the smallest periodic saw tooth lattice having only three unit
cells as depicted in Fig.~\ref{fig:lattice}(b). We choose $|t_{2}%
/t_{1}|\!=\!\sqrt{2}$ and $\mu_{1,2}\!=\!0$, so a flat band is present. In the
absence of any imperfection, the flat-band consists of a set of degenerate
localized states. The degeneracy, however, is sensitive to imperfections and
will be lifted in the presence of tiny inhomogeneity. In a real
superconducting simulator, this feature is to our advantage and one can map
out the imperfection of a simulator by inspecting spatial patterns of photon distributions.

To account for imperfections of realistic simulators, we include a small
fluctuation $\eta\! =\! |\delta t_{\alpha}/t_{\alpha}|$ of the hopping
coefficient $t_{1}$ or $t_{2}$. As estimated in section~\ref{Ununiform},
fluctuations of the hopping coefficients are about $1\%$. In the following we
define $t_{\alpha}^{>(<)}\! =\! t_{\alpha}(1\pm\eta)$ and choose $\eta=1\%$.
Imperfections will distort the flat band and favor a particular localized
wavefunction as the ground state. Fluctuations of a given simulator are the
deviations of its parameters from the averaged values. While fluctuations vary
from one sample to another, the magnitude of fluctuations in a given sample
may be treated as constant.

\subsection{Single particle picture and noninteracting bosons}

\begin{figure}[t]
\begin{center}
\includegraphics[width=0.22\textwidth]{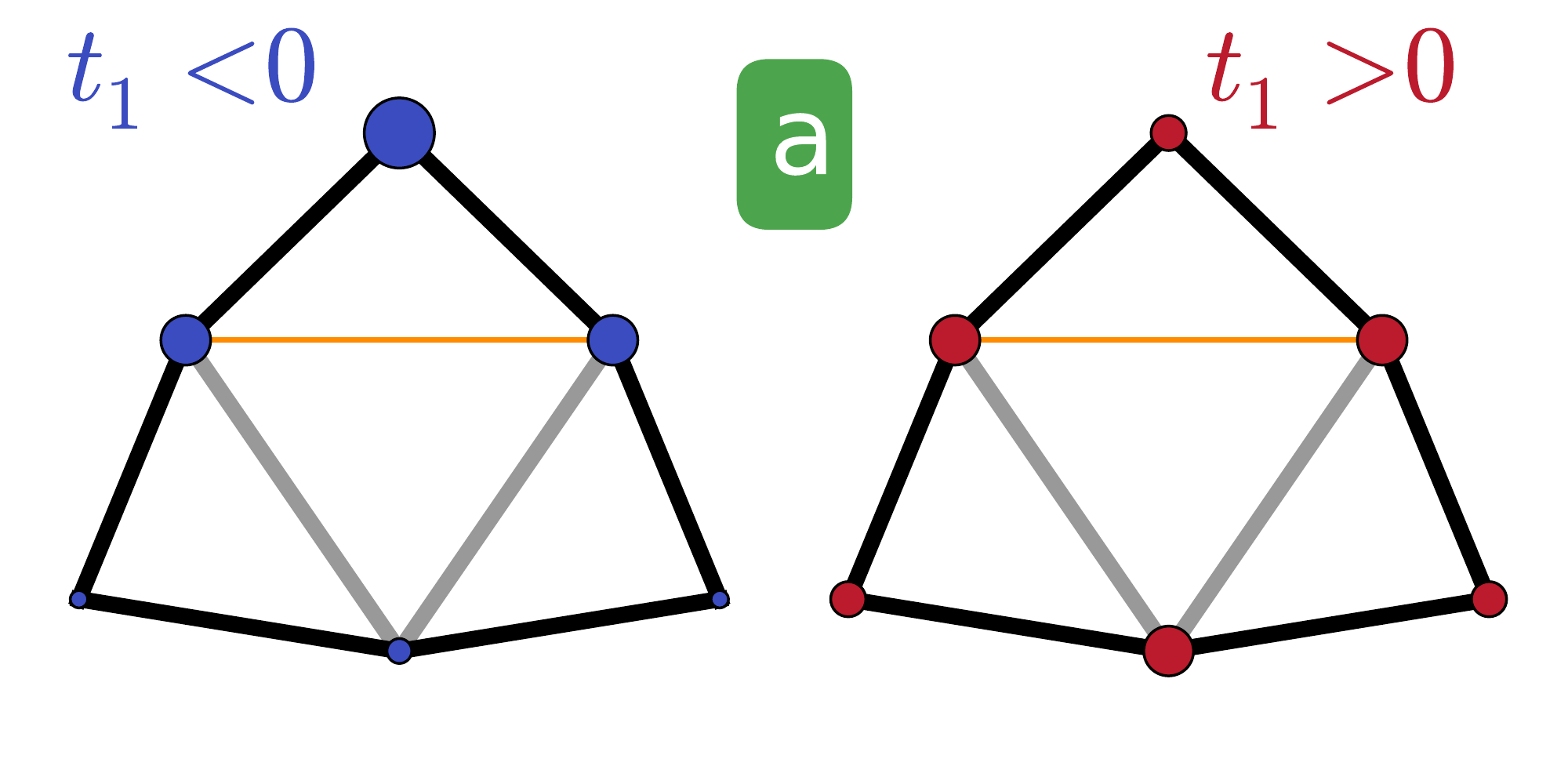}
\includegraphics[width=0.22\textwidth]{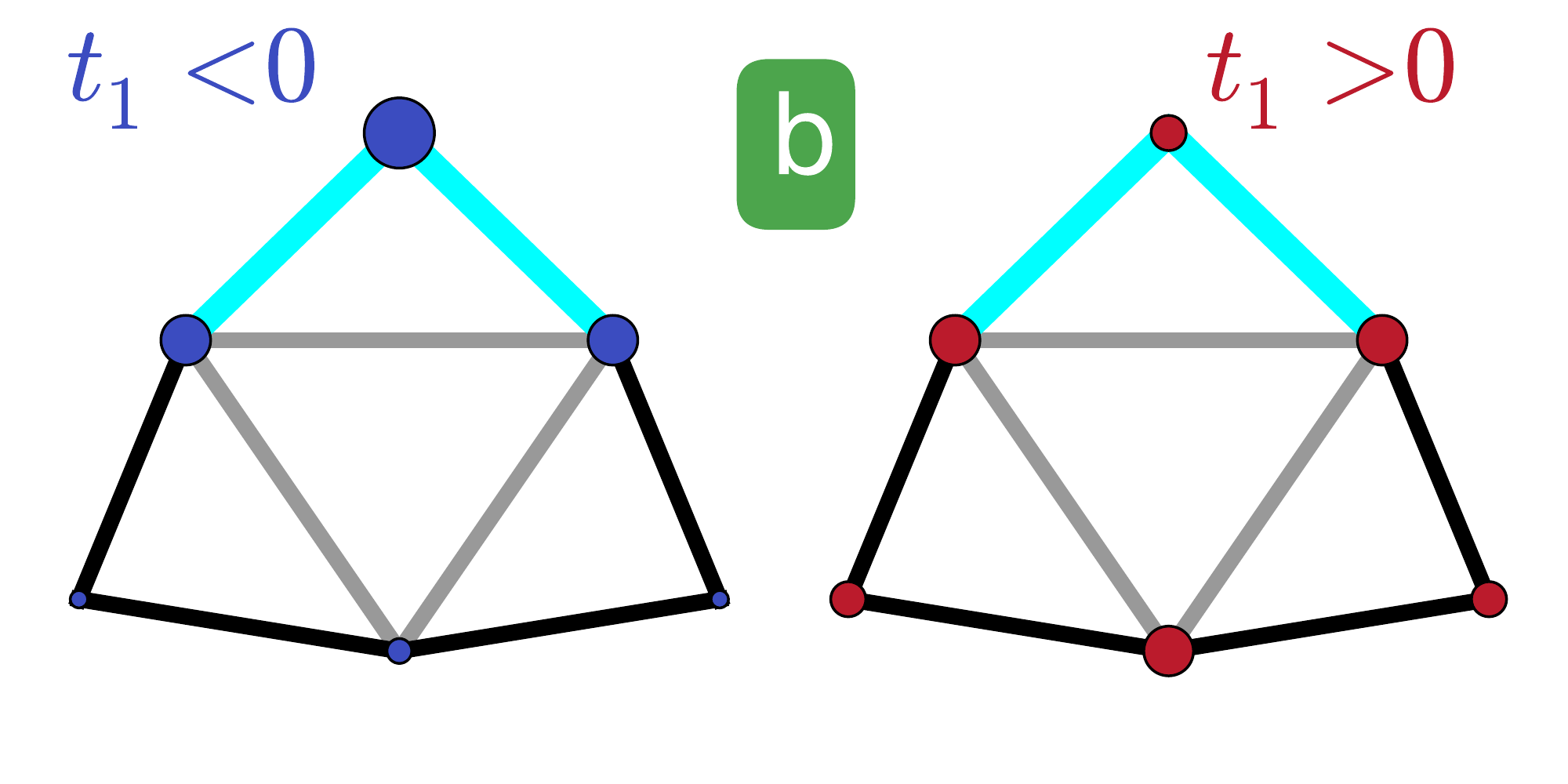}
\includegraphics[width=0.23\textwidth]{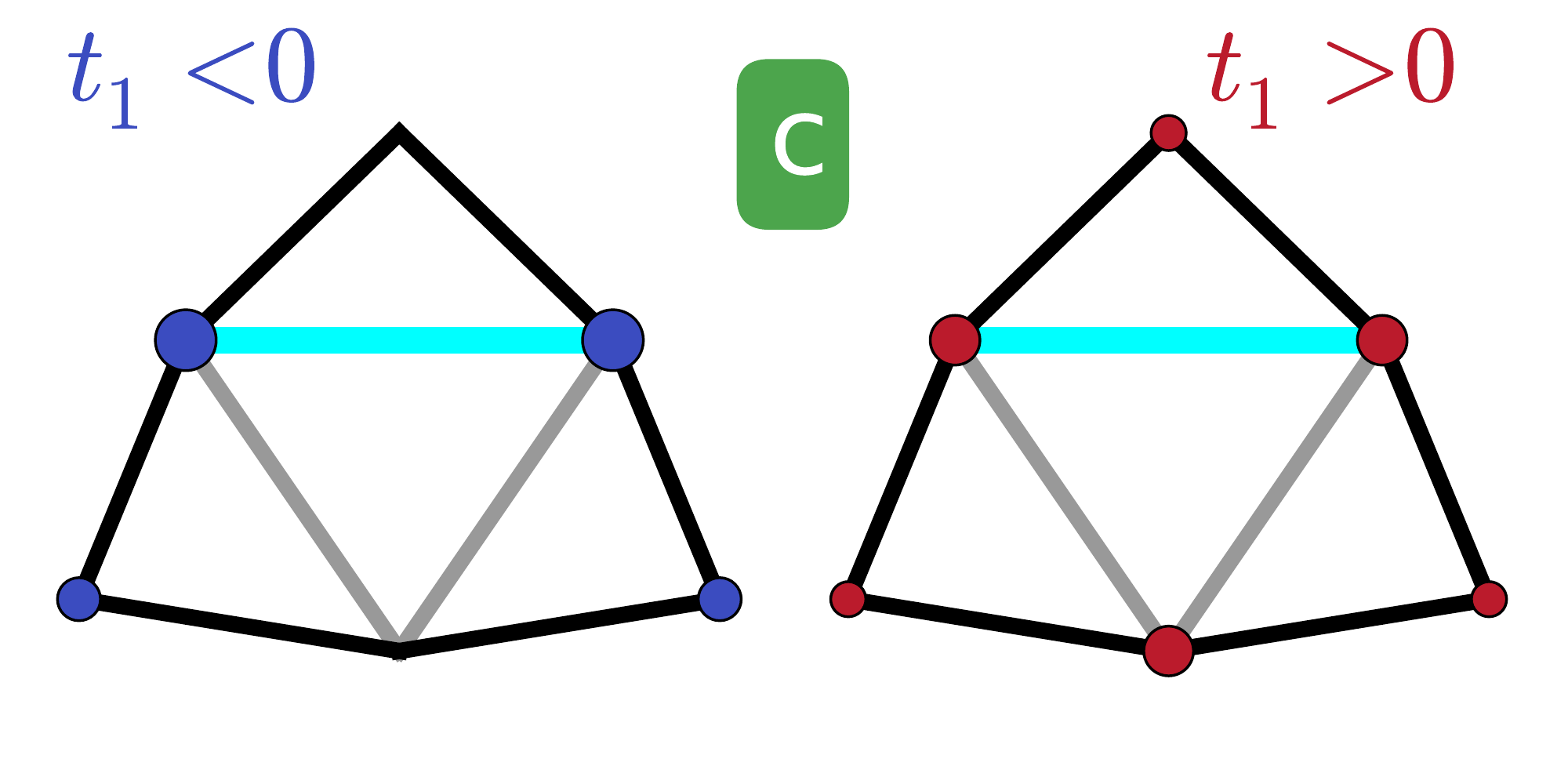}
\includegraphics[width=0.23\textwidth]{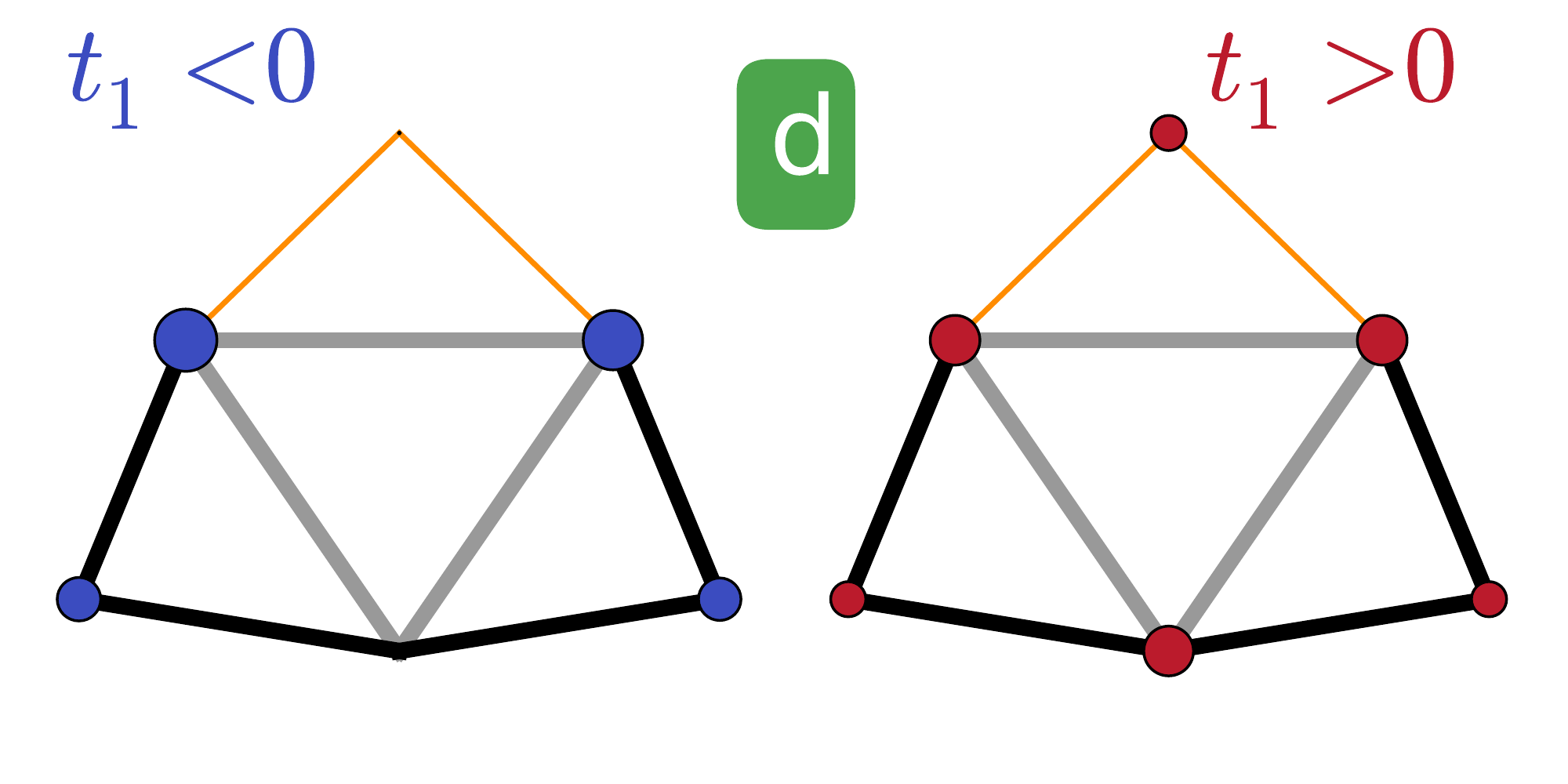}
\end{center}
\caption{(Color online) Density distributions of noninteracting particles on
periodic saw-tooth lattices with both signs of hopping coefficients,
$t_{1}\!<\! 0$ and $t_{1}\!>\! 0$. Bigger vertices represent larger particle
density, and different bond fluctuation scenarios are compared. Here
$|t_{2}/t_{1}|=\sqrt{2}$. (a) A single weaker A-A bond $t_{1}^{<}=t_{1}%
(1-\eta)$ indicated by the thin orange line. (b) A pair of stronger A-B bonds
$t_{2}^{>}=t_{2}(1+\eta)$ marked as the blue thick cyan lines. The density
pattern is similar to (a). (c) Single stronger A-A bond $t_{1}^{>}%
=t_{1}(1+\eta)$ indicated as the thick cyan line. (d) shows the dual case of
(c) with a pair of bond fluctuations $t_{2}^{<}=t_{2}(1-\eta)$. }%
\label{fig:NBDensity}%
\end{figure}

The single particle picture applies when there is only one particle in the
system and also to noninteracting bosons in the ground state. By diagonalizing
the tight-binding Hamiltonian (\ref{eq:ham}) on a finite size array, the wave
function, density distribution, and energy spectrum of the system can be
obtained. We show the main result in Fig.~\ref{fig:NBDensity}. As mention
previously, the flat band becomes the lowest lying band when $t_{1}<0$. In a
finite-size system, the flat band corresponds to several degenerate states.
Since any superposition of the degenerate states is still a valid flat-band
state, in a perfectly uniform system flat-band states could be constructed
from superpositions. In contrast, fluctuations of the parameters due to
imperfections of the simulator will lift the degeneracy and select out a
particular ground state. Mapping out the correspondence between the geometry
and the pattern of localized states then allows one to visualize features
associated with the flat band.

In Fig.~\ref{fig:NBDensity}(a), $t_{1}<0$ and the dominant fluctuation is on a
single A-A bond with $t_{1}^{<}=t_{1}(1-\eta)$, which makes the bond weaker
(the orange thin line) than the other A-A bonds (in gray). The ground state
density distribution is shown, where the sizes of vertices are proportional to
the particle density. Particles tend to accumulate on the tip of the triangle
with the weaker A-A bond. The reason for this localized state is minimization
of kinetic energy in the zero temperature limit. When $t_{1}>0$, the
dispersive band becomes the lower-energy band. Then particles will occupy the
ground state of the dispersive band, and the density distribution becomes
uniform as shown in Fig.~\ref{fig:NBDensity}(a). A localization-delocalization
transition of ground states can be observed by comparing density patterns of
the two cases with opposite signs of $t_{1}$ while choosing all A-B bonds with
$t_{2}=\sqrt{2}|t_{1}|$.

Next, we consider fluctuations of the two A-B bonds within a triangle with
$t_{2}^{>}=t_{2}(1+\eta)$ (the cyan thick lines), which are stronger than the
other A-B bonds (in black) as shown in Fig.~\ref{fig:NBDensity}(b). In this
case, localized states for $t_{1}<0$ have a similar particle distribution as
the case with a weaker A-A bond and delocalized states occur when $t_{1}>0$.
The two dual cases have similar density distributions because of similar
energy shifts in the triangle with bond fluctuations. We also consider another
dual pairs by altering the signs of the fluctuations of the previous two
cases. For a stronger A-A bond in Fig.~\ref{fig:NBDensity}(c), the density
tends to localize on both ends (blue circles) of the stronger bond (cyan thick
line) when the flat band is the lowest band ($t_{1}<0$). This case is dual to
the case with a pair of weaker A-B bonds (orange thin lines) depicted in
Fig.~\ref{fig:NBDensity}(d). By measuring the particle density of a localized
state and identifying its pattern, one can infer the location of the dominant
imperfection on a simulator.

The single-particle picture is closely related to noninteracting bosons
because each boson will occupy the single-particle ground state in the
zero-temperature limit. The ground-state density of many bosons is thus the
number of particles multiplied by the single-particle ground state density,
which can be measured in the superconducting simulator by mapping out the
photon number on each site. Thus the ground-state density of a noninteracting
bosonic system can be amplified by populating more photons in the system and
as a consequence, signatures of the localization-delocalization transition
from tuning the signs of hopping coefficients will be more prominent.

\subsection{Interacting Bosons}

\begin{figure}[t]
\begin{center}
\includegraphics[width=0.49\textwidth]{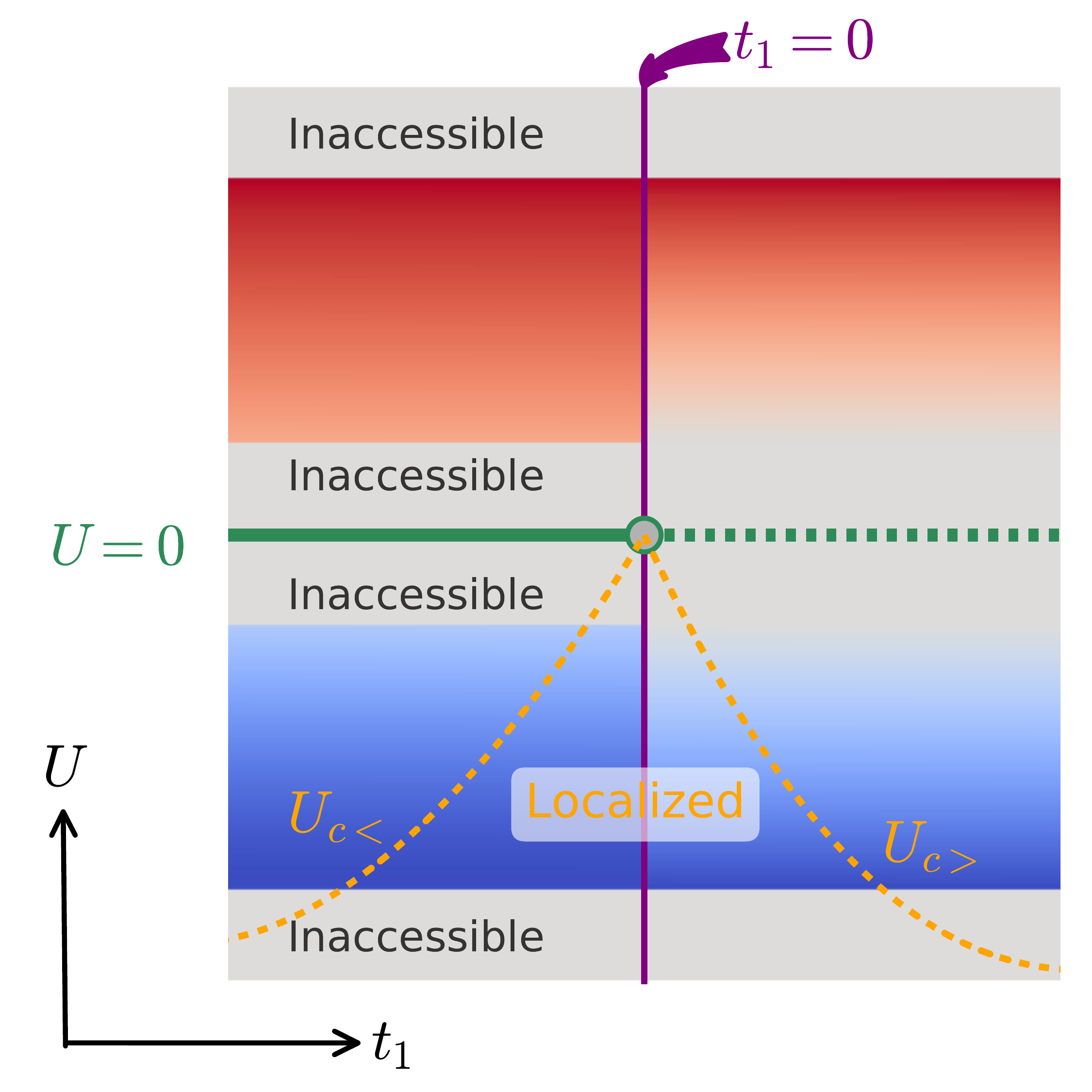}
\end{center}
\caption{(Color online) Schematic phase diagram of the proposed
superconducting circuit simulator of the BHM with two bosons on a periodic
saw-tooth lattice supporting a flat band. Imperfections of the simulator
modeled as a weaker A-A bond $t^{<}_{1}=t_{1}(1-\eta)$ have been included. The
gray areas indicated as \textit{inaccessible} are beyond the validity of the
simulator. The green solid line indicates localized states associated with the
flat band, and the green dashed line indicates delocalized states. In the
regime labeled "Localized" below the dashed yellow lines, strong attractive
interactions leads to the domination of superpositions of states with all the
particles concentrating on each site. For two particles in the system,
second-order processes favor uniformly distributed density, while for more
than two particles localized density patterns emerge.
}
\label{fig:main}%
\end{figure}

When the on-site qubit couples to the TLR in the dispersive
regime~\cite{koch2009superfluid, deng2015sitewise}, the photons acquire
effective on-site interactions and follow the BHM Hamiltonian,
Eq.~\eqref{eq:ham}. Here we focus on the case of uniform interactions where
$U_{i}=U$, $\forall i$. The interaction introduces correlations among the
bosons and invalidates the single-particle picture. We therefore use the exact
diagonalization (ED) to determine the energy spectrum and ground state
wave-function~\cite{Zhang:2010bh}. The main results are summarized in
Fig.~\ref{fig:main}, where the system has two interacting bosons in both
repulsive ($U>0$) and attractive ($U<0$) regimes. Moreover, both positive and
negative hopping coefficients ($t_{1} >0$ and $t_{1}<0$) are considered. In
the noninteracting ($U=0$) case, a localization-delocalization transition
occurs when the sign of the hopping coefficient $t_{1}$ changes.
Noninteracting localized states are labeled by the green solid line for
$t_{1}<0$ in Fig.~\ref{fig:main}. As the repulsive interaction increases, we
analyze if localized states can be stable against the on-site self-interaction
when $t_{1}\!<\!0$.

\begin{figure*}[t]
\begin{center}
\includegraphics[width=0.95\textwidth]{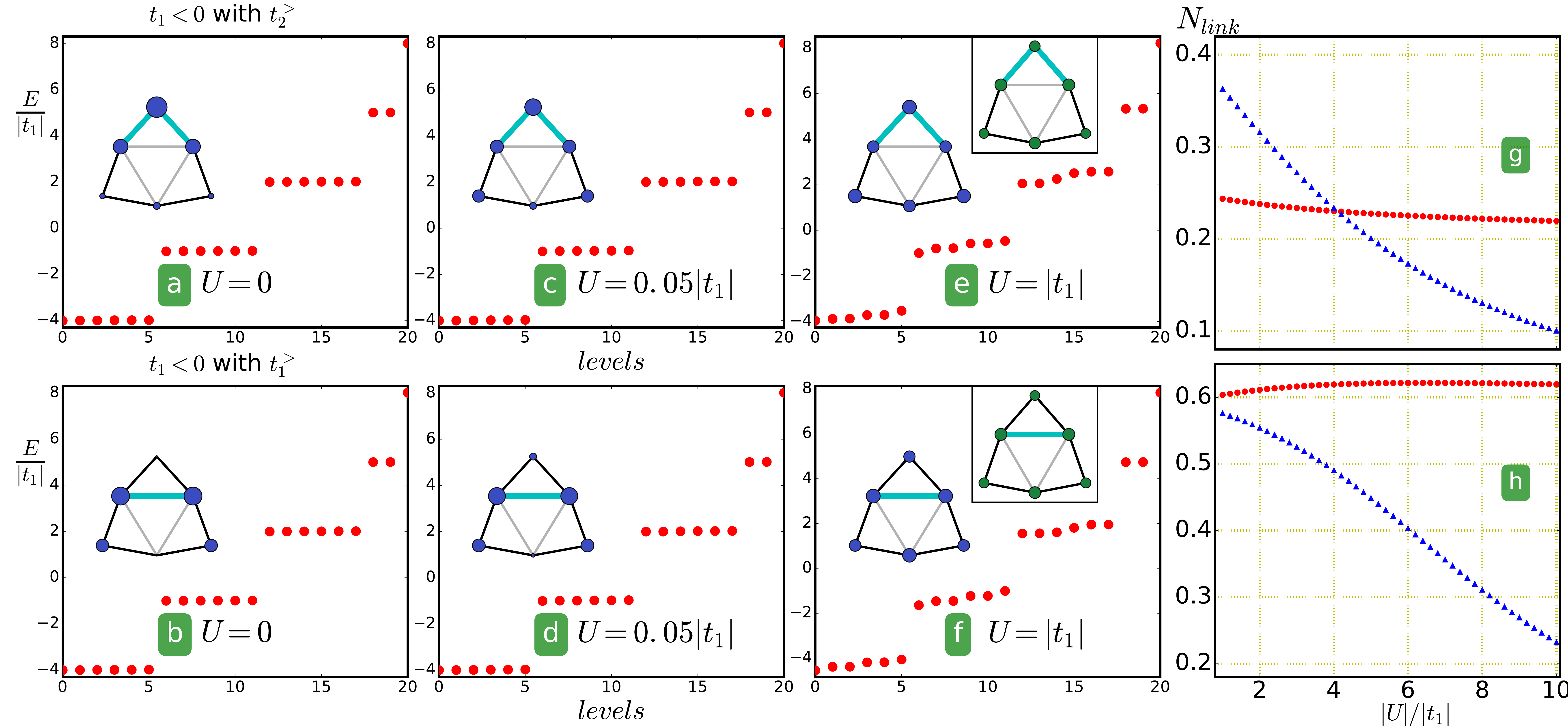}
\end{center}
\caption{(Color online) Energy levels and ground state density patterns
(inset) of two bosons with $t_{1}\!<\!0$ and different interaction strength.
Here $|t_{2}/t_{1}|=\sqrt{2}$. The left column shows a system with a pair of
bond fluctuations $t_{2}^{>}$ indicated by the thick cyan lines in the insets.
The right column shows a system with one bond fluctuation $t_{1}^{>}$
indicated by the thick cyan line in the insets. A bigger vertex indicates a
larger density in each inset. The non-interacting cases, (a) and (b), show
nearly degenerate low energy states. Imperfections of the simulator lift the
degeneracy of low-energy eigenstates and single out a unique ground state.
When $U/t_{1}$ is small, splitting of low-energy levels is tiny and not
discernible with the resolution shown here. For a small repulsive interaction
$U=0.05|t_{1}|$ in (c) and (d), we still see localized ground state density
patterns, and low-energy levels stay nearly degenerate. As the interaction
gets stronger ($U=|t_{1}|$ in (e) and (f)), density patterns become uniformly
distributed and low-energy levels are dispersive. The insets in (e) and (f)
inside the square frames show the density patterns of the attractive cases
with $U=-|t_{1}|$ and same imperfections. The density distributions are
similar. (g) DCAL $N_{link}$ (defined in Eq.~\eqref{eq:DCAL}) for $t_{1}>0$.
(h) DCAL for $t_{1}<0$. }%
\label{fig:IBDensity}%
\end{figure*}

To facilitate a fair comparison with the noninteracting case, two similar sets
of fluctuations of the parameters modeling imperfections of the simulator are
also included. The first one has a pair of stronger A-B bonds ($t^{>}_{2}$)
and the other has a single stronger A-A bond ($t^{>}_{1}$). The former is
similar to its dual case with a single weaker A-A bond ($t^{<}_{1}$) while the
latter shows similar localized wavefunctions as the case with a pair of weaker
A-B bonds ($t^{<}_{2}$). Similarities of particle-density patterns of
localized state between dual cases are still observable even in the presence
of weak self-interaction.

For repulsive interaction $U>0$, we plot the energy spectra and density
distributions for three selected values of the interaction in
Fig.~\ref{fig:IBDensity} including two sets of parameter fluctuations. When
$U=0$, the ground state is localized and the density distribution shows
localized patterns. As interaction strength increases, the density
distribution starts to spread out. In Fig.~\ref{fig:IBDensity}(a), particles
accumulate on the triangle with a pair of stronger A-B bonds when $U\! =\! 0$.
The results coincide with the single particle picture in this limit. Here
localized ground states resulting from the flat-band manifest themselves as
degenerate eigenvalues shown in the energy spectrum. We remark that, in the
presence of imperfections and interactions, the flat-band is distorted so the
ground state can be uniquely determined. As the interaction increases to
$U\!=\! 0.05|t_{1}|$ depicted in Fig.~\ref{fig:IBDensity}(c), the density
distribution of the ground state spreads out rather than localizing on a
single triangle. Before the density distribution becomes completely uniform,
particles tend to occupy the triangle on the opposite side of the strongest
fluctuation, which will be discussed later in the context of a larger system.

Finally, when the interaction strength reaches the same order as the hopping
coefficient, $U\! =\! |t_{1}|$ in Fig.~\ref{fig:IBDensity}(e), the particle
distribution becomes uniform and the flat-band spectrum no longer exists
because localized states are not eigenstates of the full Hamiltonian when the
interaction is strong. Similar phenomena are also discovered in the other case
shown in Fig.~\ref{fig:IBDensity}(b), (d), (f), where the system has only one
single stronger A-A bond. In the presence of strong self-interactions, the
ground state exhibits uniformly distributed particle density.

Similarly, the spreading of the density distribution can be observed when the
interaction is attractive. From the real-space density distribution, one
cannot discern the difference between the results from intermediate attractive
and repulsive interactions because both cases show almost uniform
distributions. However, a two-particle correlation called the density
correlation across the link (DCAL) can distinguish features of the two
interaction regimes. The DCAL is defined as
\begin{equation}
N_{link}=\sum_{<ij>}\left\langle \hat{n}_{i}\hat{n}_{j}\right\rangle .
\label{eq:DCAL}%
\end{equation}
Here the summation only includes the pairs of sites across links, $\hat{n}%
_{i}$ is the particle number operator on site $i$, and $\langle\cdots\rangle$
denotes ground-state expectation value. In the repulsive regime, the DCAL
varies slowly with the interaction and reaches a finite steady value in the
strongly interacting regime. On the other hand, the DCAL in the attractive
regime keeps decreasing. Moreover, the DCAL vanishes when the attraction
exceeds a critical value, $|U|\!>\!|U_{c<}|$. This indicates the dominance of
superpositions of doubly occupied states taking the form $\Psi=\sum_{i}%
u_{i}|0,\cdots, n_{i}\!=\!2,\cdots,0\rangle$, where the Fock states label the
particle number on each site. In this case, $N_{link}=0$. It has been proposed
that for the attractive BHM in the thermodynamic limit, the difference between
states with variable particle numbers per site and states consisting of
superpositions of fully occupied sites will become a phase
transition~\cite{buonsante2010quantum, gangat2013deterministic}, and our
results are in line with the prediction. The yellow dashed lines in
Fig.~\ref{fig:main} indicate a change of ground state properties in the
attractive regime. By measuring the numbers of photons on all sites repeatedly
and constructing their products, the DCAL can be extracted.

\begin{figure}[t]
\begin{center}
\includegraphics[width=0.48\textwidth]{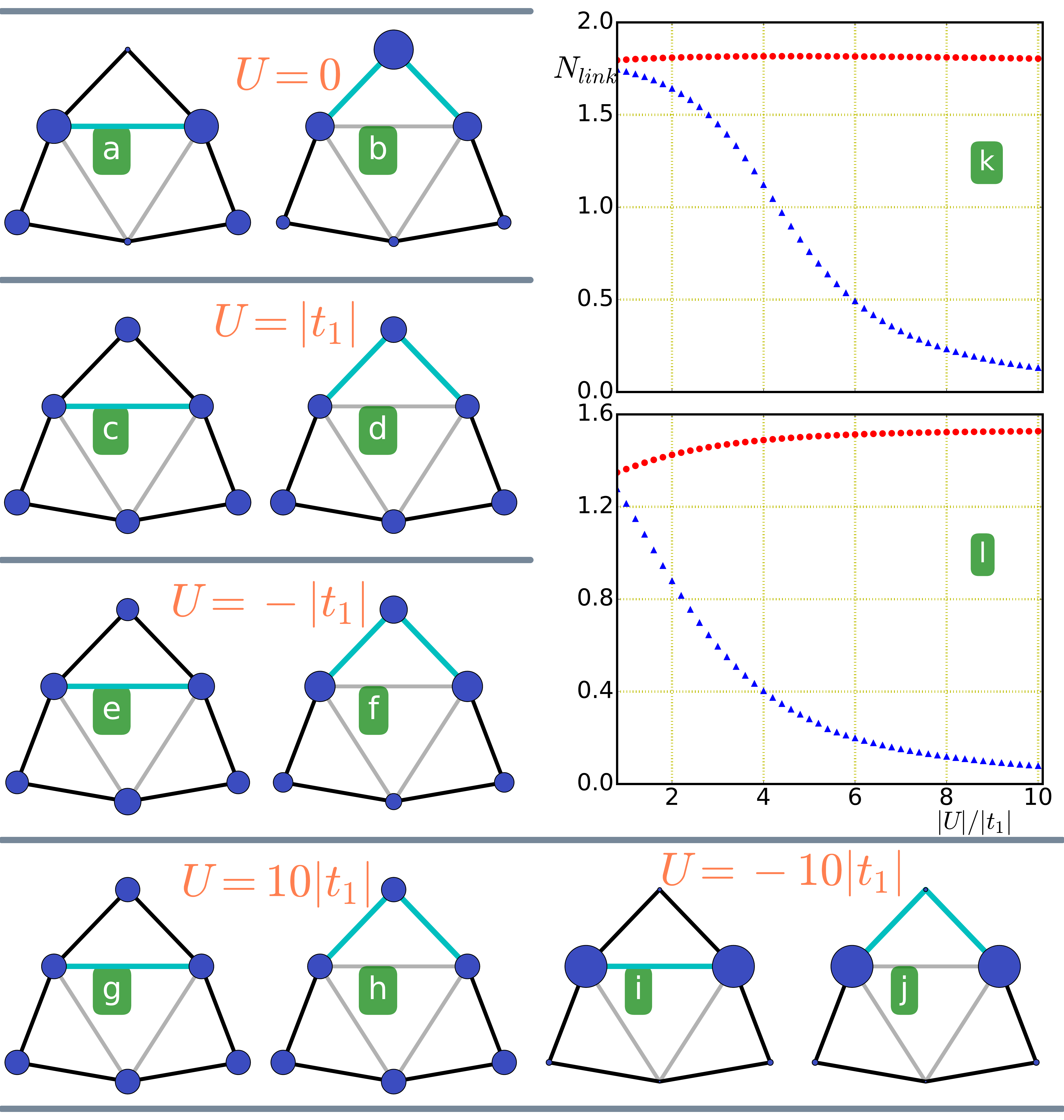}
\end{center}
\caption{(Color online) Density distributions of three bosons on a periodic
saw-tooth lattice. (a)-(b) Noninteracting cases, (c)-(d) weakly repulsive
cases, (e)-(f) weakly attractive cases, (g)-(h) strongly repulsive cases, and
(i)-(j) strongly attractive cases. Here $t_{1} < 0$ with bond fluctuations
(cyan lines) $t^{>}_{1}$ on a single A-A bond (left column) and $t^{>}_{2}$ on
a pair of A-B bonds (right column). $|t_{2}/t_{1}| = \sqrt{2}$. The localized
states in the noninteracting cases have the same structure as the case with
two particles. In the repulsive regime, the density patterns gradually become
uniformly distributed. In contrast, another regime with localized density
patterns emerges in the strongly attractive regimes. The DCAL $N_{link}$
(defined in Eq.~\eqref{eq:DCAL}) for $t_{1}>0$ and $t_{1}<0$ are shown in (k)
and (l), respectively. The repulsive case (red circles) and attractive case
(blue triangles) approach different values in the strongly interacting
regimes.
}
\label{fig:N3}%
\end{figure}

When more than two particles are loaded into the system, the
localization-delocalization transition at $U=0$ is still observable as the
hopping coefficient $t_{1}$ changes sign. Fig.~\ref{fig:N3} shows the case of
three bosons on a periodic saw-tooth lattice. The spreading trend of the
density at intermediate $U$ is also similar to the two-particle case.
A difference between $N=2$ and $N>2$, where $N$ is the
total number of bosons in the system, is that localized density patterns emerge again in
the strongly attractive regimes ($U<0$ and $|U/t_{1}|>>1$) as shown in
Fig.~\ref{fig:N3}(i) and (j). The reason for the re-entrance of localized
density patterns can be understood from second-order degenerate perturbation
theory and the presence of imperfections of the system parameters. Second-order hopping processes
select out sectors in the Fock space consisting of states like $|0,\cdots
,0,N,0,\cdots,0\rangle$ according to the inhomogeneity of the hopping coefficients, which then cause concentration or distillation of the density
in the region with the strongest fluctuations. Interestingly, when there are
only two bosons in the system, the selection process favors uniform density distribution. Moreover, the localized density patterns in the
strongly attractive regime for $N>2$ can be observed with negative as well as positive
hopping coefficients, but the localized density patterns are different when
the hopping coefficients change sign. A more rigorous analysis of a three-site
system is given in Appendix~\ref{app2}. No re-entrance of localized density patterns
are observed in the repulsive regime for $N=2,3$ and this should also apply to
$N>3$.

The DCAL in the strongly repulsive regime saturates to a value depending on
the total particle number, but in the attractive case the DCAL always decays
toward zero as shown in Fig.~\ref{fig:N3} (k) and (l). Therefore, the DCAL
reveals the subtle difference between states in the attractive and repulsive
regimes accessible by the superconducting circuit simulator with or without
inhomogeneity in the parameters. We remark that weakening of localized states
by interactions in the BHM has been proposed in Refs.~\onlinecite{fisher1989boson,buonsante2004topology}.

\subsection{Larger arrays of the simulator}

\begin{figure}[t]
\begin{center}
\includegraphics[width=0.43\textwidth]{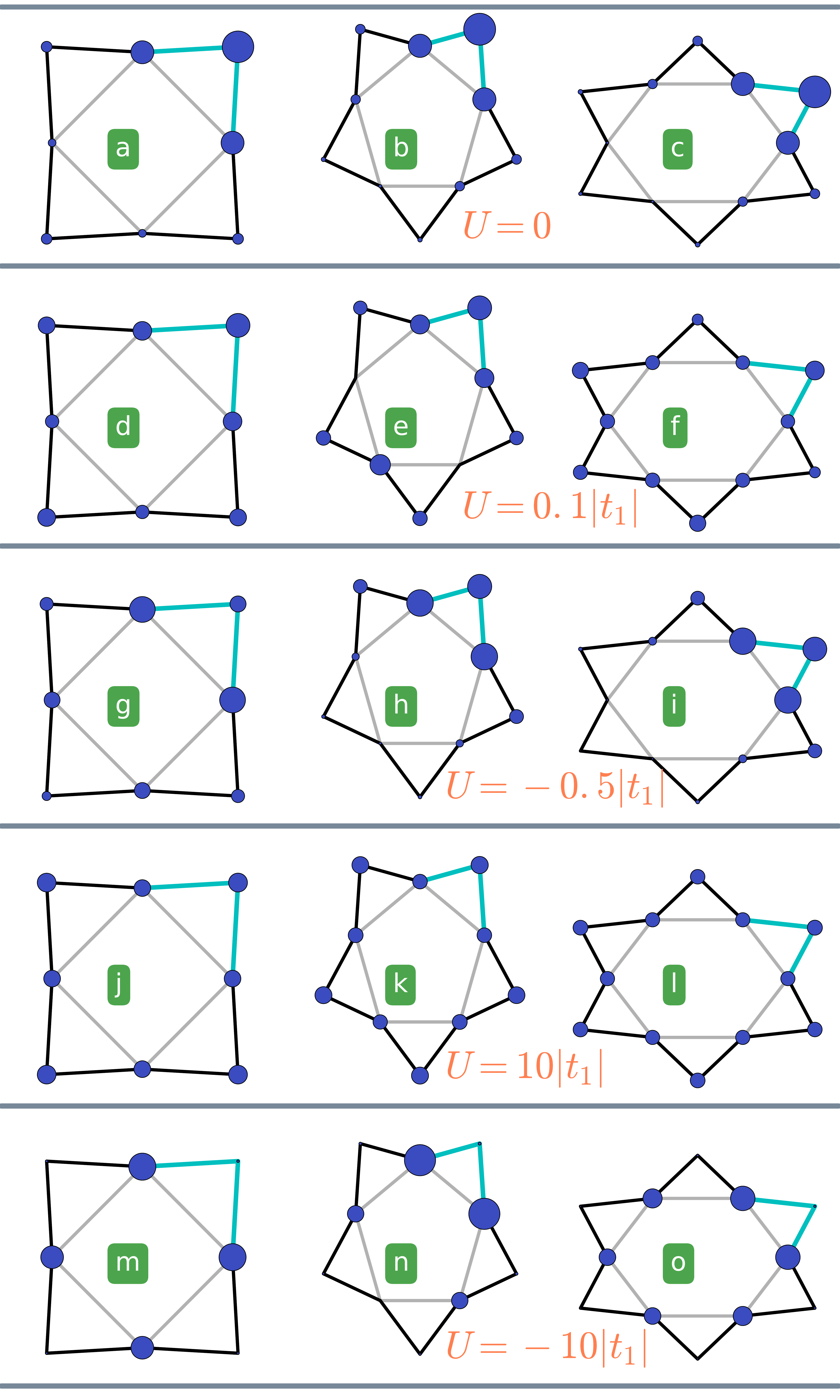}
\end{center}
\caption{(Color online) Density distributions of three bosons on
different lattices: Four unit cells in the left column, five unit cells in the
middle column, and six unit cells in the right column. Here $t_{1}\!<\!0$ with
bond fluctuations $t_{2}^{>}$ on a pair of A-B bonds (cyan lines) and
$|t_{2}/t_{1}|=\sqrt{2}$. The interaction strength are: (a)-(c) $U\!=\!0$
(non-interacting), (d)-(f) $U\!=\!0.5|t_{1}|$ (repulsive), (g)-(i)
$U\!=\!-0.5|t_{1}|$ (attractive), (j)-(l) $U\!=\!10|t_{1}|$ (repulsive),
(m)-(o) $U\!=\!-10|t_{1}|$ (attractive). Bigger vertices correspond to larger
densities. The spreading to the triangle opposite to the triangle with
dominant bond fluctuations in the presence of repulsion can be clearly
observed by comparing the first and second rows. For attractive interactions,
the first and third rows show the spreading occurs near the triangle with
dominant bond fluctuations, while the last row shows the re-entrance of
localized density patterns. For a system with only two bosons, the density
pattern remains uniform in the strongly attractive regime. }%
\label{fig:IBDensityLarge}%
\end{figure}

By considering systems with four, five, and six unit cells of the saw-tooth
lattice, we can further address the behavior of localized states. The density
distributions of three bosons without interaction ($U=0$) and hopping
coefficient $t_{1}<0$ are shown in Fig.~\ref{fig:IBDensityLarge}(a)-(c) with
different numbers of unit cells respectively. As mentioned before, localized
states are favorable on the triangle with the strongest bond fluctuations due
to imperfections of the simulator. Fluctuations of the hopping coefficient
with realistic parameters can be controlled to be within $1\%$.

In the weakly repulsive regime, for instance $U=0.1|t_{1}|$ shown in
Fig.~\ref{fig:IBDensityLarge}(d)-(f), the particles tend to occupy the
opposite side of the triangle with the most fluctuations in order to stay away
from each other. This tendency avoids the on-site self-interaction already at
the zeroth order. On the other hand, for the attractive cases shown in
Fig.~\ref{fig:IBDensityLarge}(g)-(i) we also see spreading of localized states
with $U=-0.5|t_{1}|$. Although the density patterns in both strongly
attractive and repulsive regimes are similar, the trends are different. In the
attractive regime the density starts to populate the region near the triangle
with the largest bond fluctuations rather than occupying the opposite triangle
in the repulsive cases.

As the repulsive interaction gets stronger, for example $U=10|t_{1}|$
shown in Fig.~\ref{fig:IBDensityLarge}(j)-(l), the density distributions
eventually become uniform. On the other hand, strongly attractive interaction
drives the system with $N>2$ into another regime with localized density patterns as shown in Fig.~\ref{fig:IBDensityLarge}(m)-(o) with $U=-10|t_{1}|$. For $N=2$, the density distribution remains uniform in the strongly attractive regime. The
strongly attractive regime with $t_{1}>0$ or $t_{1}<0$ are labeled as
"Localized" in Fig. \ref{fig:main} to emphasize the emergence of localized
density patterns when more than two interacting bosons are present.

\subsection{Implications for experimental implementations}

Before closing our discussion, we briefly comment on experimental realizations
of the simulator.

\emph{State preparation:} Innovative ways have been investigated for preparing
a deterministic product Fock state in multi-TLR
systems~\cite{underwood2012low, mariantoni2011photon, houck2012chip}. For
instance, in Ref.~\onlinecite{deng2015sitewise} the TLRs in the simulator (the
on-site TLR) may be connected to auxiliary control TLRs via additional tunable
couplers which can be formed by several SQUIDs. It has been assumed that one
can generate any numbers of photon in the control
TLR~\cite{wallraff2004strong}. By adjusting the qubit energy and sending in
pump signals to compensate for the detuning between the on-site TLR and the
control TLR, one can swap the photon states between the on-site TLR and
control TLR~\cite{sharypov2012parametric} and prepare a deterministic Fock
state in the on-site TLR. In this scheme, the numbers of photons on different
sites can be different. A product Fock state can be prepared as the initial
state of the simulator by applying this process on all the sites simultaneously.

The measurement of the photon number on each site may be performed in a
similar way to map out the state in Fock space~\cite{deng2015sitewise}.
Alternatively, quantum non-demolition (QND) measurements may be performed on
the on-site TLRs so the photon numbers on each TLR could be measured with a
weakly-coupled and off-resonant probe signal and the photons in the simulator
will not be demolished \cite{johnson2010quantum,
guerlin2007progressive,schuster2007resolving}.

\emph{Cooling:} Localized versus delocalized ground states of photons have
been contrasted in Figs.~\ref{fig:NBDensity} to~\ref{fig:IBDensity}. Once the
simulator is prepared in a product Fock state, photons are either localized on
certain sites or spreading among different sites. In order to study ground
state properties, one needs to cool the system down to the ground state
without losing the photons in the system. Hence, a cooling scheme needs: i)
Collective excitations of the simulator need to be reduced while conserving
the total photon number, which could be achieved by using a cavity-assisted
side-band cooling scheme. An experiment has demonstrated a collective ground
state of the BHM in the attractive regime simulated by an array of
transmons~\cite{hacohen2015cooling}. There have been other theoretical
studies~\cite{mirrahimi2014dynamically,
sarlette2011stabilization,roy2015continuous} on number-conserving
manipulations of the photon excitations in superconducting circuits providing
feasible alternatives. ii) The cooling rate, defined as the sum of the
stimulated (pump-assisted) transmission from higher levels to the ground
state~\cite{hacohen2015cooling}, of the whole lattice has to be faster than
the decay rate of photons. Considering the high Q nature of the TLR, the
lifetime of photons can be around milliseconds~\cite{wang2009improving,
reagor2013reaching,rigetti2012superconducting}. Thus the cooling rate needs to
be in the range of kHz to MHz.

\emph{Tunability of parameters:} Broad tunability of the simulator is made
possible by the following mechanisms. i) The qubit coupled to the TLR
introduces an effective tunable on-site interaction. The repulsive or
attractive interaction depends on whether the qubit is red-detuned or
blue-detuned from the resonator, and the interaction strength depends on the
magnitude of the detuning. ii) The SQUIDs coupling neighbor sites allow the
hopping coefficient to be tuned from negative values to positive values. The
Josephson junction in a SQUID is modeled as a combination of a capacitor and a
tunable nonlinear inductor \cite{MakhlinRevModPhys}. When the inductor
dominates, the SQUID coupler of the simulator becomes a low pass filter so
photons, which are AC electromagnetic signals in the TLR, tend to be blocked
by the inductor and yields a positive hopping coefficient. When the capacitor
dominates the coupling, the coupler becomes a high pass filter. Therefore,
photons tend to hop between different sites and reduce overall energy by
hopping, which yields a negative hopping coefficient. Combining those features
allows the proposed superconducting system to simulate various phases of the
Bose Hubbard model.

\section{Conclusion}

\label{sec:conclusion} Simulations of the BHM on periodic saw-tooth lattices
supporting a flat band are feasible by using the versatile superconducting
circuit simulator with broadly tunable parameters discussed here. A
localization-delocalization transition of noninteracting bosons associated
with the flat band of saw-tooth lattices is made possible because the sign of
hopping coefficients in the simulator can be controlled. In the presence of
onsite interactions, density patterns from localized states are still
observable and sensitive to inhomogeneity of the underlying elements. One may
exploit this feature and use the density pattern as a diagnosis tool for
identifying imperfections.

The rich phase diagram of the BHM with a flat band illustrated in
Fig.~\ref{fig:main} elucidates interesting interplays between geometry and
interaction. Delicate differences between ground states in repulsive and
attractive regimes, although not visible in the density distribution, can be
discerned by two-particle correlations. Moreover, this work may inspire
similar studies in superconducting circuit simulators of fermionic
systems~\cite{greif2013short, schneider2012fermionic, Barends:2015tv}.

\acknowledgments
We thank S. Hacohen-Gourgy and Chunjing Jia for inspiring discussions in the
early stage of this work.

\appendix

\section{Tunability and applicability of superconducting circuit}

\label{app1} Here we summarize the modeling, approximation, and experimental
parameters of the superconducting circuit simulator. The on-site term
describes a TLR coupled to a superconducting qubit in Eq.~\eqref{eq:originsc}
as
\begin{equation}
\mathcal{H}_{i}^{site} =\omega_{i}^{r}b_{i}^{\dagger}b_{i}+\frac{\omega
_{i}^{q}}{2}\sigma_{i}^{z}+g_{i}^{q}\sigma_{i}^{x}(b_{i}^{\dagger}+b_{i})
\end{equation}
with the TLR frequency $\omega^{r}\! =\! \frac{2\pi}{\sqrt{C^{r}L^{r}}}\!
=\!2\pi\sqrt{E_{C}^{r}E_{L}^{r}}$, qubit frequency $\omega^{q}\! =\!
2E_{J}^{q}\cos(\frac{\phi_{e}}{2})$ when the qubit is capacitively biased at
the charge degeneracy
point~\cite{nakamura1999coherent,clarke2008superconducting}, and the coupling
between the TLR and qubit $g_{i}^{q}=2e\frac{C_{g}} {C_{\Sigma}^{q} }%
\sqrt{\omega^{r}C^{r}}$. The Pauli operators $\sigma^{\{x,y,z\}}$ represent
the qubit coupled to each TLR. $C^{r}$ and $L^{r}$ are the total capacitance
and inductance of the TLR. $E_{C}^{r}=\frac{(2e)^{2}}{C^{r}}$ and $E_{L}^{r}
=\frac{1}{L^{r}(2e)^{2}}$ are capacitive and inductive energies of the TLR.
$E_{J}^{q}$ is the Josephson energy of each junction in the qubit. $C_{g}$ is
the coupling capacitance and $C_{\Sigma}^{q}$ is the sum of the total
effective capacitance between the TLR and the ground. $\phi_{e}$ is the
magnetic flux through the SQUID loop. After the rotating-wave approximation,
the on-site interaction coupling constant in Eq.~\eqref{Honsite} is given by
$U^{p}_{i}=g_{i}^{q}(\frac{g_{i}^{q}}{\Delta})^{3}$, which can be attractive
or repulsive by varying the detuning $\Delta=\omega^{r}-\omega^{q}$ to be
negative or positive.

The coupler SQUID in Eq.~\ref{eq:rwalink}, $\mathcal{V}_{ij}^{couple}$,
contains the capacity coupling constant $g^{cap}=G_{0}\omega^{r}E_{C}%
^{jj}/E_{c}^{r}$ and inductive coupling constant $g^{ind}=4G_{0}\omega
^{r}E_{J}^{jj}\cos(\frac{\phi_{e}}{2})/E_{L}^{r}$. $E_{J}^{jj}$ and
$E_{C}^{jj}$ are the inductive and capacitive energy of the Josephson junction
in the SQUID coupler. The amplitude factor $G_{0}$ depends on where on the TLR
the qubit is coupled to. If it is placed at the
antinode~\cite{blais2004cavity}, the factor $G_{0}=1$. In
Fig.\ref{fig:lattice} (d), the two SQUIDs below the middle of the TLR are
placed at $3/8$ and $5/8$ of the TLR, hence $G_{0}=\cos(\pi/4)$ for them. By
tuning $\phi_{e}$ in the domain $[0,2\pi]$, $g^{ind}$ can be tuned positive or
negative. The superconducting circuit model is derived in the large dispersive
regime, where $\Delta_{i}\gg g_{i}^{q}\gg t_{i}$.

We now turn to experimental parameters. Available experimental
data~\cite{wallraff2004strong,
blais2004cavity,hacohen2015cooling,lang2011observation,
hoffman2011dispersive,Baust:2014uq,mariantoni2011photon, wulschner2015tunable}
allow one to set $\omega^{r}=5GHz$, $E_{C}^{c}=0.2GHz$, $E_{J}^{c}=10GHz$,
$E_{L}^{r}=50GHz$, $E_{c}^{r}=0.5GHz$, $E_{C}^{c}=0.2GHz$, $g^{ind}\in
\lbrack-1.2,1.2]GHz$, $g^{cap}=0.6GHz$, then the hopping coefficient
$t_{i}=g^{cap}+g^{ind}\in\lbrack-0.3,1.5]GHz$. We consider an appropriate
range of $\phi_{e}$, so $t_{i}$ can be tuned in the range $[-0.6,0.6]GHz$.
Notice that $g^{cap}$ needs to be smaller than the tunable inductive term
$g^{ind}$ so the hopping coefficient can be switched between positive and
negative values. A small $E_{C}^{c}$ can be achieved by coupling capacitors
between the SQUID and TLR. Using only capacitors~\cite{houck2012chip} or
SQUIDs~\cite{wulschner2015tunable} coupled to TLRs has been demonstrated in
experiments. Here we consider a tunable coupler from a combination of a
capacitor and a SQUID in order to access both regimes with positive and
negative hopping coefficients. The effective capacitance between two TLRs
coupled by a combined coupler can be increased and a low capacitive energy
$E_{C}^{c}=(2e)^{2}/2C$ can be achieved.

Similar to the SQUID coupler, a flux bias through the qubit can be used to
tune the qubit frequency. In order to apply the simulator to the BHM,
additional approximation conditions are imposed: (a) The atomic limit
$g_{i}^{q}\gg t_{i}$, where photon hopping can be treated as a
perturbation~\cite{koch2009superfluid}. (b) The dispersive condition
$\Delta_{i}\gg g_{i}^{q}$, which decouples the qubit from the TLR and allows a
perturbative treatment of the on-site Hamiltonian and the effective on-site
interaction $U$~\cite{blais2004cavity,koch2009superfluid,houck2012chip}. (c)
The perturbative condition, which requires the hopping coefficient $t_{i}$ to
be about the same order of or smaller than the on-site interaction $g_{i}%
^{q}(\frac{g_{i}^{q}}{\Delta_{i}})^{3}$ as the TLR is coupled to the qubit.
Then perturbation theory is applicable~\cite{deng2015sitewise}.

The capacitive coupling between the qubit and resonator, $g_{i}^{q}$, is fixed
once elements are fabricated. Here we take a typical experimental value
$g_{i}^{q}=130$ MHz~\cite{wallraff2004strong, blais2004cavity}. In order to
keep the simulator in the dispersive regime, the detuning $\left\vert
\Delta_{i}\right\vert $ should be tuned within $[0.4,5)$%
GHz~\cite{blais2004cavity, hacohen2015cooling}. Therefore, the on-site
interaction $U_{i}\in\lbrack-5,0)\cup(0,5]$MHz and $t_{i}$ can be tuned within
$[-10,10]$MHz. We remark that the on-site coupling constant $U_{i}=g_{i}%
^{q}(\frac{g_{i}^{q}}{\Delta})^{3}$ can be attractive or repulsive depending
on whether the qubit is in the red-detuned ($\Delta<0$) or blue-detuned
($\Delta>0$) regime. By increasing the detuning $\Delta$, $|U_{i}|$ can be
reduced. One may expect when the qubit is far off-resonant from the TLR,
$U_{i}$ can approach $0$. However, a continuous scanning control of the flux
bias used to tune the SQUID frequency does not change the sign of $U_{i}$.
Moreover, in the presence of the qubit there is always a non-vanishing value
of $U_{i}$. To circumvent difficulties of reducing $U$ towards $0$, one may
detach the qubit from the resonator to shut down photon interactions in the
resonator completely and reach the limit $U_{i}=0$. Recent
experiments~\cite{Baust:2014uq, wulschner2015tunable} have shown ultrastrong
tunable coupling between two TLRs and they support the estimation of hopping
coefficients used here. On the other hand, in the ultrastrong coupling regime,
different nonlinear
effects~\cite{Baust:2014uq,bourassa2012josephson,wulschner2015tunable} other
than simple photon hopping could arise and validity for the BHM simulator
breaks down. To explore the phase diagram of the BHM with a flat band,
parameters of the simulator should remain in the weak coupling regime, where
coupling between neighboring sites (photon hopping) is weaker compared to the
on-site qubit-TLR coupling.

\section{Second-order degenerate perturbation in the strongly attractive
regime}

\label{app2} According to Appendix \ref{app1}, the hopping coefficients $t_{1,2}$ of the simulator can be continuously tuned around
$t_{1,2}=0$, which enables us to study the strongly interaction regime $|U/t_{1,2}|>>1$ by tuning the ratio. In the strongly
attractive regimes, a regime with localized density patterns emerges when the
system has more than two particles. Here we use second-order
degenerate perturbation theory~\cite{schiff1968quantum} to explain the
emergence of the regime and why it is only observable for $N>2$. Here we consider
the BHM on a three-site lattice forming a triangle. The Hamiltonian is
\begin{align}
H  & =H_{0}+H_{I}\nonumber\\
H_{0}  & =\frac{U}{2}\sum_{i}n_{i}(n_{i}-1)\label{aeq:h0}\\
H_{I}  & =-\sum_{\langle i,j\rangle}t_{ij}(b_{i}^{\dagger}b_{j}%
+h.c.)\label{aeq:hi}%
\end{align}
with onsite coupling constant $U$
and hopping coefficient $t_{ij}$ with $i,j$ denoting the two sites connected by a bond. In the strongly attractive regime with  $|U/t_{ij}|>>1$, we treat $H_{I}$ as a perturbation.
When all the hopping coefficients vanish, the unperturbed ground
state is any superposition of the three Fock states $|1\rangle=|N,0,0\rangle$,
$|2\rangle=|0,N,0\rangle$, and $|3\rangle=|0,0,N\rangle$ because of the attractive interaction. Thus, we take the space spanned
by the three states and consider corrections due
small $t_{ij}$. The unperturbed Hamiltonian
is degenerate in the $\{|1\rangle,|2\rangle,|3\rangle\}$ basis and the matrix representation takes the form
\begin{equation}
H_{0}=\left(
\begin{array}
[c]{ccc}%
E_{N} & 0 & 0\\
0 & E_{N} & 0\\
0 & 0 & E_{N}%
\end{array}
\right)
\end{equation}
with $E_{N}=UN(N-1)/2$. For the three degenerate unperturbed states, all
first-order terms vanish because they cannot be connected by exchanging only
one particle.

The hopping terms, however, introduce second-order processes with the
assistance from higher-energy unperturbed states like $|N-1,1,0\rangle$, etc.
For instance, $\Delta E_{11}=\sum_{f}\langle N,0,0|H_{I}|f\rangle\langle
f|H_{I}|N,0,0\rangle/(E_{f}-E_{N})$, where the intermediate states are
$f\in\{|N-1,1,0\rangle$, $|N-1,0,1\rangle\}$ with energy
$E_{f}=U(N-1)(N-2)/2$. The ground state can then be found by diagonalizing the
matrix~\cite{schiff1968quantum}
\begin{equation}
\left(
\begin{array}
[c]{ccc}%
\Delta E_{11} & \Delta E_{12} & \Delta E_{13}\\
\Delta E_{21} & \Delta E_{22} & \Delta E_{23}\\
\Delta E_{31} & \Delta E_{32} & \Delta E_{33}%
\end{array}
\right)  .\label{eq:SecondOrderMatrix}%
\end{equation}
In constructing $\Delta E_{ij}$, the unperturbed states are excluded from
being used as intermediate states. The diagonal terms $\Delta E_{jj}$ are of
the order of $|t_{ij}^{2}/U|$. The off-diagonal terms, however, are sensitive
to the total particle number $N$. When $N=2$, one can see that two degenerate
unperturbed states can be connected via an intermediate higher-energy state.
For instance, $|2,0,0\rangle$ can hop to $|1,1,0\rangle$ and then to
$|0,2,0\rangle$. Therefore, all $\Delta E_{ij}$ are of the order of
$|t_{ij}^{2}/U|$ and the second-order ground state after diagonalizing the
matrix is still a superposition of the three unperturbed ground states. As a
consequence, the density is mostly uniform.

In contrast, two different unperturbed states cannot be connected via
second-order processes when $N>2$. For example, when $N=3$, there is no
intermediate state $|f\rangle$ connecting $|3,0,0\rangle$ and $|0,3,0\rangle$
with two hopping events. Therefore, $\Delta E_{ij}=0$ if $i\neq j$ at the second-order level. The
diagonal terms, however, are finite at the second-order level. Moreover, in
the presence of fluctuations of the hopping coefficients, $t_{ij}$ are
different and this leads to different $\Delta E_{jj}$. Therefore, the matrix
\eqref{eq:SecondOrderMatrix} picks up a preferred state in its
diagonalization. For example, if $\Delta E_{11}$ is the smallest among $\Delta
E_{jj}$, the ground state up to the second order would be $|N,0,0\rangle$.
When $N$ and the number of sites are large, the ground state may remain degenerate up to the second order in a
subspace of the original set of unperturbed states, and higher-order
perturbations will further lift the degeneracy. The important point is that
the ground state, up to the second order, only includes a subset of the
unperturbed states, which means the density is concentrated or distilled on
certain sites. Therefore, the ground state in the presence of weak hopping
coefficients and imperfections of the parameters exhibits localized density patterns when $N\ge 3$.

When there are more than three sites, the second-order degenerate perturbation
theory still applies and one expects localized density patterns in the
strongly attractive regime when $N>2$. Moreover, the perturbation theory works
for both negative as well as positive $t_{ij}$. The localization patterns,
though, are different when the sign of $t_{ij}$ changes because higher-order
processes sensitive to the sign will further refine the selection of the
ground state. When the interaction is repulsive, the ground state always tends
to spread out the density and no localization is found in the strongly
repulsive regime.


%

\end{document}